\documentclass[conference]{IEEEtran}
\IEEEoverridecommandlockouts
\usepackage{cite}
\usepackage{amsmath,amssymb,amsfonts}
\usepackage{algorithmic}
\usepackage{graphicx}
\usepackage{textcomp}
\usepackage{xcolor}
\usepackage{subcaption}
\usepackage{comment}
\usepackage[font=small,labelfont=bf]{caption} 

\def\BibTeX{{\rm B\kern-.05em{\sc i\kern-.025em b}\kern-.08em
    T\kern-.1667em\lower.7ex\hbox{E}\kern-.125emX}}
\begin{document}

\title{Development of a fabrication-to-benchtop process for SiN-based quantum devices\\
{\footnotesize \textsuperscript{*}}
\thanks{National Research Council of Canada, National Science and Engineering Research Council of Canada, Canadian Foundation of Innovation.}
}

\author{\IEEEauthorblockN{1\textsuperscript{st} Connor Kupchak}
\IEEEauthorblockA{\textit{Department of Electronics} \\
\textit{Carleton University}\\
Ottawa, Canada \\
connor.kupchak@carleton.ca}
\and
\IEEEauthorblockN{2\textsuperscript{nd} Abubaker Tareki}
\IEEEauthorblockA{\textit{Department of Electronics} \\
\textit{Carleton University}\\
Ottawa, Canada \\
abubakertareki@cunet.carleton.ca}
\and

\IEEEauthorblockN{3\textsuperscript{nd} Tara Moradi}
\IEEEauthorblockA{\textit{Department of Electronics} \\
\textit{Carleton University}\\
Ottawa, Canada \\
tmoradi@gmail.com}
\and
\IEEEauthorblockN{4\textsuperscript{rd} Patrick	Laferriere}
\IEEEauthorblockA{\textit{Department of Electronics} \\
\textit{Carleton University}\\
Ottawa, Canada \\
patricklaferriere@cunet.carleton.ca}
\and
\IEEEauthorblockN{5\textsuperscript{th} Niall Tait}
\IEEEauthorblockA{\textit{Department of Electronics} \\
\textit{Carleton University}\\
Ottawa, Canada \\
nialltait@cunet.carleton.ca}
\and
\IEEEauthorblockN{6\textsuperscript{th} Khaled Mnaymneh}
\IEEEauthorblockA{\textit{Quantum and Nanotechnologies Research Centre} \\
\textit{National Research Council of Canada}\\
Ottawa, Canada \\
khaled.mnaymneh@nrc-cnrc.gc.ca}

}

\maketitle

\begin{abstract}

Silicon nitride has emerged as a key material for the advancement of future quantum communication and computational devices. However, challenges persist in establishing a comprehensive development process, spanning from fabrication to optical characterization, essential for advancing these devices to their next stage of potential. Maintaining control over the entire process, from start to finish, alongside device modeling, enables more efficient and timely iteration and optimization.  Here we introduce a fabrication-to-benchtop process. We employ an in-house stoichiometric recipe for device fabrication and conduct comprehensive characterization of optical losses and material properties. Understanding and controlling the refractive index of the device are vital for predicting device geometries and group indices. Implementing the refractive indices of our fabricated devices, we simulate microring resonators and demonstrate their application in generating optical solitons.
\end{abstract}

\begin{IEEEkeywords}
Integrated Photonics,
Quantum Optics,
Chip Fabrication,
Device Modeling
\end{IEEEkeywords}

\section{Introduction}

Silicon nitride (SiN) is a unique material for fabricating compact, integrated photonic devices with a wide range of applications. Its attractive properties include inherently low losses, a favorable refractive index, and a wideband transparency window that spans from the UV to mid-infrared (mid-IR) regions \cite{IEEE_1}. These attributes minimize detrimental two-photon absorption and scattering effects, making SiN ideal for high-energy applications. Additionally, SiN has a high nonlinear index (nearly 10 times that of silica), enabling low-power on-chip optical processing for nonlinear effects such as four-wave mixing. These characteristics, combined with its compatibility with metal-oxide-semiconductor (CMOS) processes, make silicon nitride a sought-after material for integrated optical devices and a solution to the complexities associated with optical fibre systems.  In classical communications, SiN has demonstrated its maturity in applications such as information processing \cite{SiNapps}, optical switching \cite{SiNswitch}, microwave photonics \cite{SiNmicrowave}, nonlinear conversion \cite{SiNfreqconver}, and (de)multiplexing in the O-band wavelength window \cite{ObandDemux}.

In novel quantum devices, SiN has been used for on-chip single-photon sources \cite{SiNsouce,SiSource2}, time-bin encodings \cite{SiNtimebin,SiNtimebin2}, squeezed light generation \cite{SiNsqueeze1,SiNsqueeze2,OnChipSqueezing}, and single-photon detection \cite{SiNdetector}. The potential for low-energy operations is advantageous for quantum optical information processing, facilitating applications such as quantum simulation \cite{QuantumSim} and quantum machine learning \cite{QuantumML}. Here, photons are the information carriers, and operations are achieved with large-scale interferometers that critically depend on controlling the optical path length and phase. Integrated photonics is key to harnessing phase control \cite{PhaseStable}. With precise phase control, more advanced quantum devices can be designed, including those with reconfigurability using splitters and shifters \cite{ChipReconfig}, balanced homodyne detection \cite{ChipBHD}, and continuous variable quantum information processing \cite{CVQKD}.

The addition of SiN devices expands the existing on-chip quantum optical network toolbox, which has been utilized in applications such as teleportation \cite{chipteleport}, key distribution \cite{chipQKD}, and photonic gates \cite{PhysRevLett.128.060501}. To achieve the full potential of these applications, it is essential to understand and control the SiN fabrication process, as well as to dictate the nominal loss and dispersion properties of the devices.

In this work, we present progress toward a comprehensive process spanning from fabrication-to-benchtop testing. We begin with an in-house stoichiometric recipe and proceed with a deposition procedure for device fabrication. By employing various optical characterization techniques, we quantify the optical losses and refractive index behavior of our fabricated devices. With  knowledge of the devices material properties, we perform simulations of the optical processes inside microring resonators. These simulations demonstrate how complete process control can aid in the optimization of waveguide geometries for the production of optical solitons. This work may provide a fabrication-to-benchtop blueprint for transformative advancements in SiN-based quantum communication devices.

\section{Deposition}

\begin{figure}[htbp]
\centerline{\includegraphics[width=0.4\textwidth]{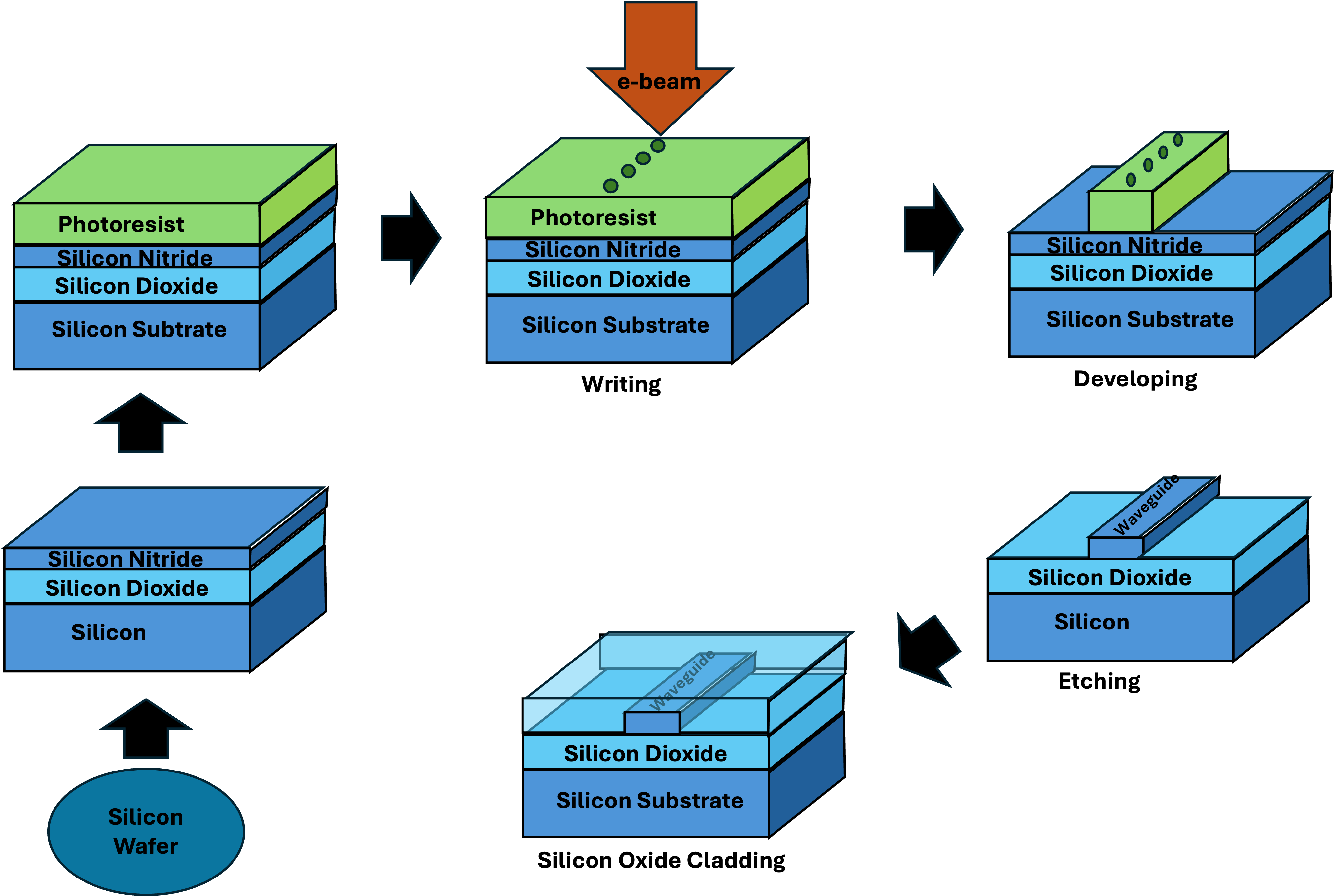}}
\caption{Schematic diagram of LPCVD backetch deposition process.}
\label{figDep}
\end{figure}

The most ideal silicon nitride films have a stoichiometric composition with a silicon-to-nitride ratio of 0.75 \cite{IEEE_stoich}. The nominal refractive index for these films is just under~$n=2$; values above 2 indicate a higher silicon content, which is associated with increased optical losses.

Several fabrication techniques are compatible with CMOS processes, including low-pressure chemical vapor deposition (LPCVD) and plasma-enhanced chemical vapor deposition (PECVD). In general, LPCVD produces higher-quality films than PECVD and is often the preferred deposition method. However, LPCVD has limitations due to the high temperatures and low pressures required to achieve the desired low optical losses and high nonlinear indices in silicon nitride. Additionally, LPCVD can result in high film stresses that may cause film cracking, rendering them unsuitable for device applications \cite{IEEE_crack}. To address these issues, more advanced techniques like the Damascene process are used to reduce film stress.  In this work, we employ a back-cutting technique with the LPCVD process for deposition. A schematic diagram of our procedure is shown in Fig. \ref{figDep}.

The higher temperatures associated with LPCVD result in better film quality due to fewer impurities and lower associated loss compared to PECVD. However, LPCVD is a slower process and creates larger stress values in the material than PECVD \cite{IEEE_ratio}. During LPCVD, the thermal energy is high enough to break H-bonds and remove impurities. The resulting loss values are strongly correlated with the N/Si ratio in the films. Specifically, the propagation losses at 1310 nm can decrease to $<$1.0 dB/cm with an increasing N/Si ratio. Si-rich films lead to more Si-H and Si-Si bond defects, which cause larger propagation losses. In general, SiN films exhibit higher losses at 1550 nm compared to 1310 nm due to the presence of N-H bonds in the films. However, note that in the telecommunication C-band, SiN exhibits negligible two-photon absorption at 1550 nm \cite{No2photon}.

For our deposition process, we used two gases: dichlorosilane (SiH$_2$Cl$_2$) and ammonia (NH$_3$). The ratio of DCS/NH$_3$ determines the stoichiometric properties of the films. A ratio of $<1$ will result in nitride-rich films that exhibit less loss than silicon-rich films. The refractive index of SiN can be tuned between 1.7 and 3.0 by adjusting the N/Si ratio of the material. The backside nitride removal was performed using reactive ion etching (Vision RIE).

Our designed deposition process allows precise control over the silicon nitride thickness, enabling effective dispersion engineering and the fabrication of compact device footprints. For novel quantum optics applications, we require a thickness greater than 500 nm \cite{SiN_thick}. As an intermediate step, we are presenting film depositions between 200-300 nm thick, with SiN deposited on silicon wafers and a 5 $\mu m$ thermal oxide layer used as an isolation layer.

To alleviate the high stress associated with the LPCVD technique, we implemented several mitigation strategies. These strategies included scribing lines to section off crack-free areas \cite{IEEE_scribe} and using the Damascene process \cite{IEEE_dama}. During this process, we monitored the refractive index values for changes caused by different stress levels resulting from these mitigation techniques.

\section{Characterization techniques}

To ensure film quality after deposition, we implemented the following techniques to assess film thickness, composition, stress, and optical loss:

\begin{itemize}
\item Thickness and refractive index: A sufficient device thickness is necessary for quantum optical applications, and optimal losses require the device refractive index to be near that of stoichiometric silicon nitride. In our process chain, the thickness and refractive index of the material were quantified using an ellipsometer (J.A. Woollam M2000). The thickness was further verified by scanning the films with an electron microscope. Additionally, the stoichiometric nature of the films was confirmed through Tetra-methyl ammonium hydroxide (TMAH) etching.

\item Chemical composition and impurities: The chemical composition was analyzed using X-ray photoelectron spectroscopy (XPS), and the presence of impurities in the film was assessed via Fourier transform IR spectroscopy. Impurities can be identified by the bonding signatures of hydrogen with other elements in the film.

\item Stress measurement: A film stress measurement tool (FSM 900TC) was used to determine wafer bow and quantify film stress at room temperature. Stress values were evaluated based on the wafer bow exhibited before and after nitride backside processing.

\item Optical losses: Optical losses experienced during optical propagation through the fabricated film were quantified using a Metricon model 2010/M prism-coupler.
\end{itemize}

A prior study showing a more detailed review our SiN characterization techniques can be found in Ref. \cite{Tareki2023}. Here, we fabricated 260 nm thick stoichiometric waveguides using LPCVD at 820$^\circ$C and 300 mTorr. The waveguides were deposited on 2-inch silicon wafers, which were then cleaved using a thermal oxide isolation layer. We used Si$_3$N$_4$ cutback methods for etching.

\begin{figure}
\centering
\begin{subfigure}[b]{0.5\textwidth}
   \includegraphics[width=1\linewidth]{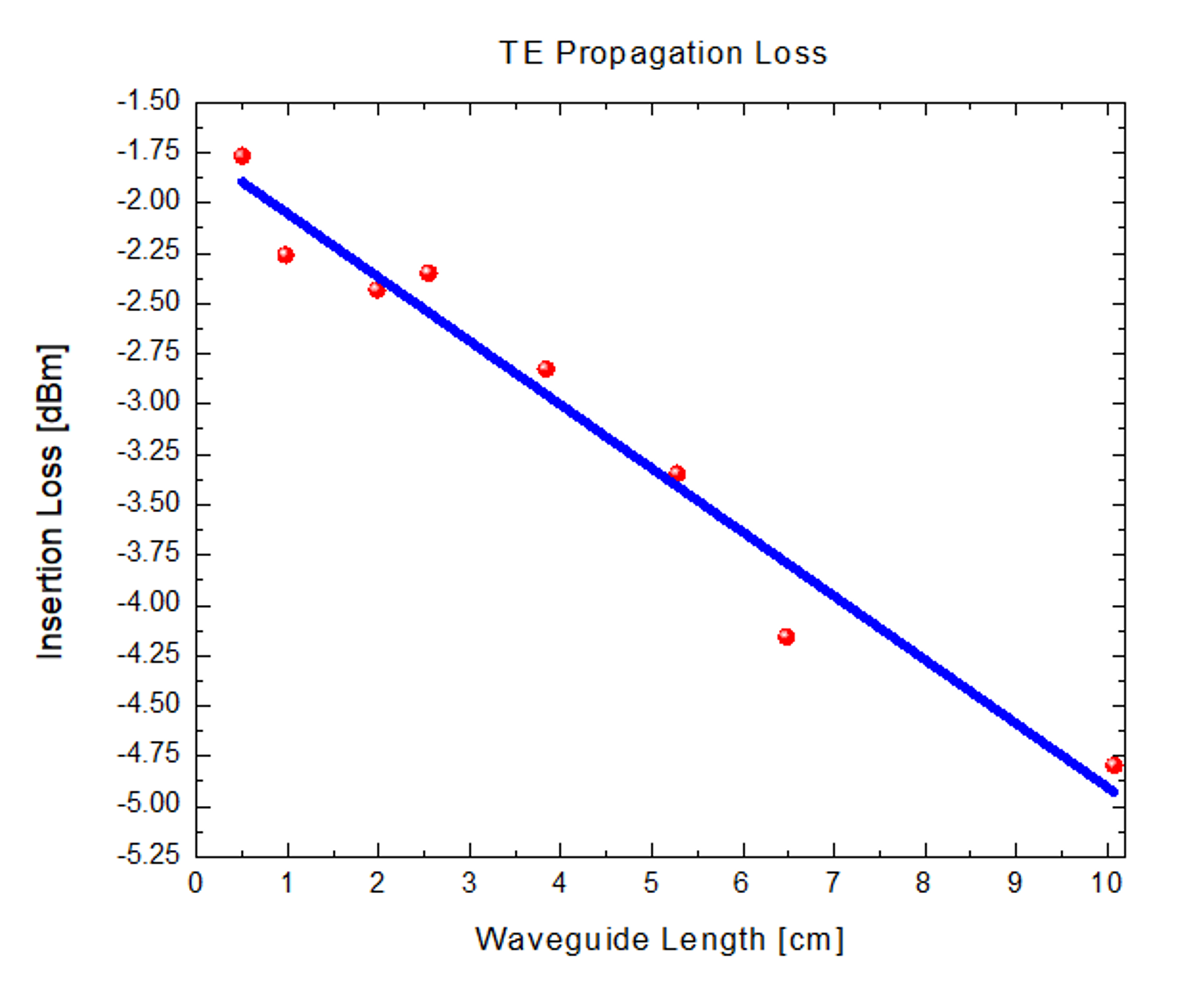}
   \caption{}
   \label{fig:Ng1} 
\end{subfigure}

\begin{subfigure}[b]{0.5\textwidth}
   \includegraphics[width=1\linewidth]{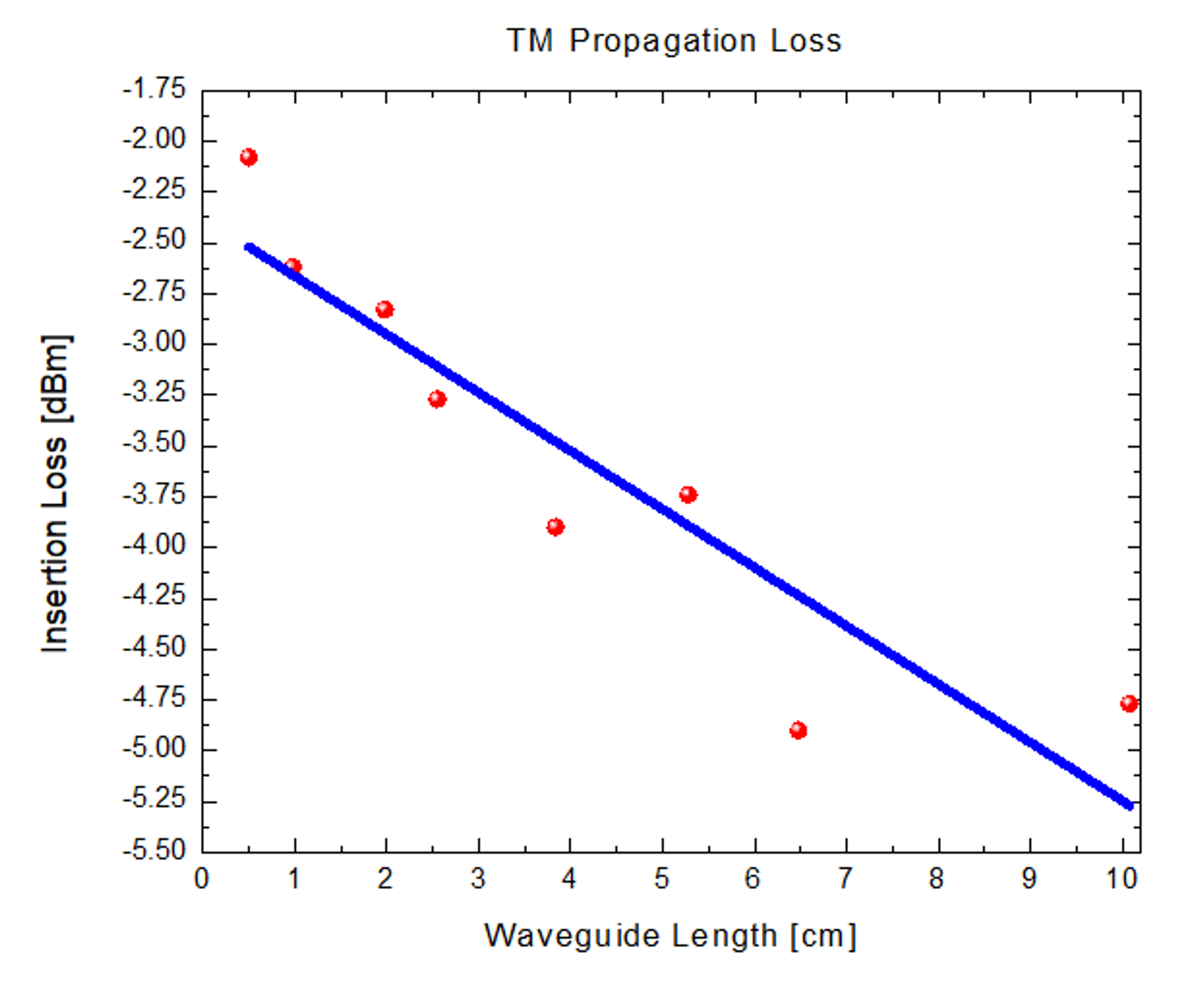}
   \caption{}
   \label{fig:Ng2}
\end{subfigure}

\caption[Optical Loss characterization]{Optical losses of the silicon nitride waveguides fabricated with our LPCVD deposition process. Loss measurements were done at 980 nm for the (a) TE and (b) TM modes.}
\end{figure}

For the initial optical loss characterization, we employed a NIR laser at 980 nm. Light was coupled into the input and output facets of the device using lensed fibres. Figures \ref{fig:Ng1} and \ref{fig:Ng2} display the loss characterization for TE and TM modes of the planar waveguides, respectively. The in-house fabricated devices showed a propagation loss of 0.3 dB/cm and a coupling loss of 1 dB per facet. These values are comparable to those achieved in industrial labs.

\section{Applications to microring resonators}

\subsection{General device modeling}

\begin{figure}
\centering
\begin{subfigure}[b]{0.4\textwidth}
   \includegraphics[width=1\linewidth]{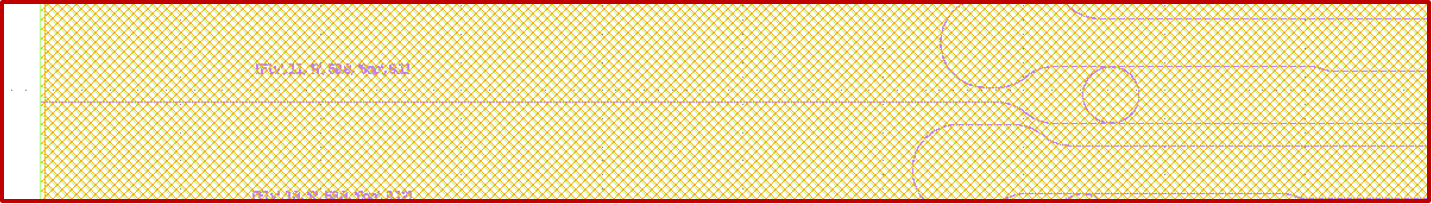}
   \caption{}
   \label{fig:RR1} 
\end{subfigure}

\begin{subfigure}[b]{0.4\textwidth}
   \includegraphics[width=1\linewidth]{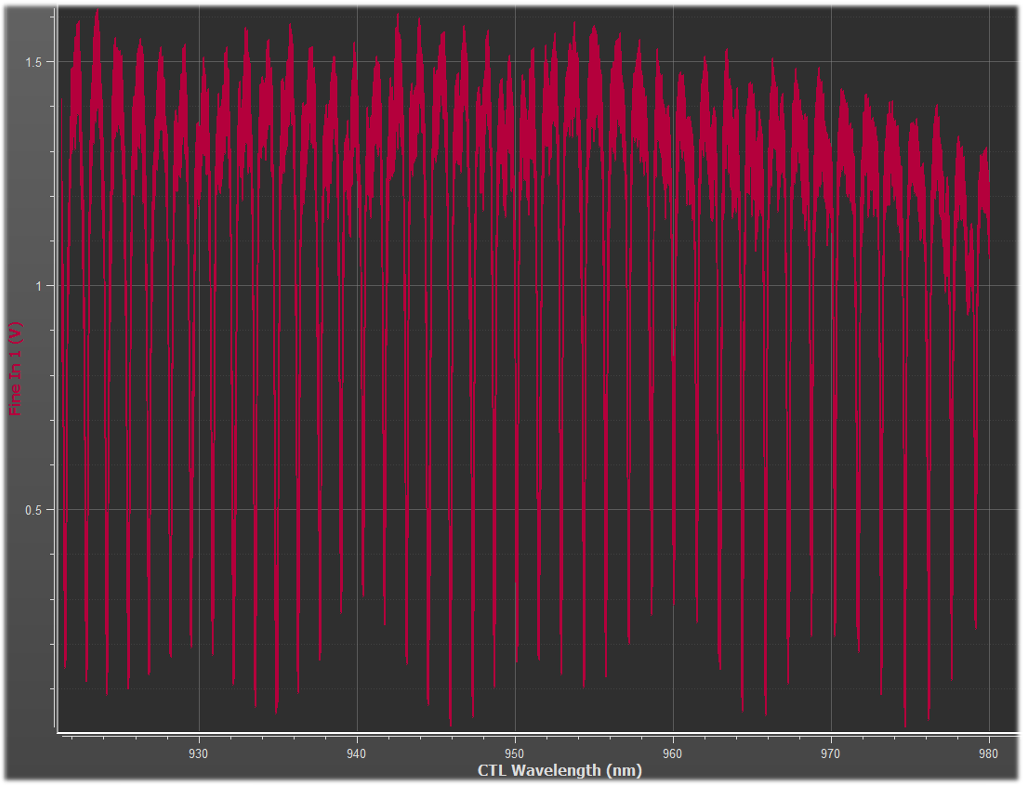}
   \caption{}
   \label{fig:RR2}
\end{subfigure}

\caption[Ring Resonator design on SiN]{(a) Two-track ring resonator design for generating optical frequency combs and solitons. (b) Resonance behavior of a on-chip ring resonator device when the Wavelength is scanned.}
\end{figure}

High-Q microring resonators are an attractive device for SiN fabrication due to their many potential applications. Microring resonators made from high $\chi^{(3)}$ materials exhibit Kerr nonlinearities that are ideal for parametric frequency conversion and four-wave mixing processes. This has garnered interest in using microring resonators for generating optical frequency combs and optical solitons \cite{ChipSoliton}. An example of a SiN microring resonator fabricated using our in-house deposition process is shown in Fig. \ref{fig:RR1}.

Optical frequency combs and solitons have advanced applications in sensitive frequency measurements, precision spectroscopy, astronomy, ultra-short pulses, and complex electromagnetic waves. In telecommunications, frequency combs may offer a solution for increasing bandwidth, with each "tooth" of the comb encoding a specific information bit. Additionally, frequency combs are being explored for quantum communication applications, including quantum key distribution \cite{GuoQKD}.

Fabricating a resonator for a specific application requires precise device modeling and simulation beforehand, as the device behavior can be sensitive to the exact waveguide geometries. Many frequency-dependent properties, such as dispersion and group index, can be verified through resonance characterization of the microring. A typical resonance scan for an in-house fabricated SiN microring resonator is shown in Fig. \ref{fig:RR2}.

We are particularly interested in designing and simulating microring resonators for optical Kerr solitons that operate in the telecommunications O-band. Access to the O-band is crucial for extending information capacity beyond the current telecommunications C-band. In our modeling, we focus on ensuring sufficient light confinement by choosing a high contrast between the core and cladding refractive indices. This approach enhances light confinement, reduces the device footprint, allows for precise dispersion engineering, and provides improved nonlinear interactions at lower optical powers.

The physics of microring resonators can be simulated using either coupled mode equations (a frequency-domain approach) or the Lugiato-Lefever equation (LLE), a time-domain approach that utilizes the split-step Fourier method (SSFM). We choose the LLE approach here due to its computational efficiency in calculating a large number of modes. The general LLE equation is as follows:

\begin{equation}\label{NSE11}
  \begin{aligned}
  t_{R} \dfrac{\partial A}{dt}=\sqrt{\theta}A_{in}+
     &(\alpha^{'}+i\delta_{0})A+\\
     & L \Bigg(i\sum_{k \geqslant 2}\dfrac{\beta_{k}}{k!}(i\dfrac{\partial}{\partial{T}})^{k}A+i\gamma |A|^{2}{A}\Bigg).
 \end{aligned}
\end{equation}

Here, $t_R$ is the resonator’s round-trip time; $A$ is the field amplitude inside the ring, which depends on both the slow time $t$ and the fast time $T$; $\theta$ is the power transmission coefficient; $A_{in}$ is the input pump field; $\alpha$ represents the losses; $\delta_0$ is the cavity detuning; $L$ is the cavity length; $\beta_2$ is the second-order dispersion coefficient; $\gamma$ is the nonlinear coefficient; and $k$ indexes the dispersion orders. The first term on the right-hand side of the LLE represents the external input, while the subsequent terms represent linear losses, detuning, dispersion, and the Kerr effect, respectively.
 
When simulating the physics of a microring resonator, the first step is to perform a linear analysis to calculate the resonant frequencies and the corresponding transmission and reflection spectra. This linear analysis can be conducted using various methods, such as the transfer matrix method, the finite-difference time domain (FDTD) method, or the finite-element method (FEM).  The microring resonator under study is of the add-through type, where a continuous wave input is applied at one end of the bus waveguide (the add port) and combs are produced at the other end (the through port). The dimensions of the ring, including the cross-sectional area and radius, were chosen to achieve the anomalous dispersion necessary for frequency-comb generation. This will be discussed in more detail in the next section.

\subsection{Microring dispersion modeling}

Dispersion is a crucial property in microring resonator modeling, as the frequency dependence of the refractive index determines phase velocities and optical path lengths, which are key to comb generation. In the absence of dispersion and non-linearity, ring resonators will only exhibit naturally standing wave resonances (modes). The central mode occurs at frequency $\omega_0$, with side modes $m$ occurring at $\omega_{m} = \omega_0 \pm m \times D_1$, where $m$ is the mode number and $D_1$ is the first-order dispersion, which is characterized by the free spectral range (FSR) of the resonator ($D_1 = \text{FSR}$).  When the model is extended to include higher-order dispersion parameters, such as second-order ($D_2$), third-order ($D_3$), and fourth-order ($D_4$) dispersion, their impact on the microdisk resonances must be considered. It is these higher-order dispersion terms that are responsible for phenomena like optical soliton generation.

\begin{figure}[htbp]
\centerline{\includegraphics[width=0.5\textwidth]{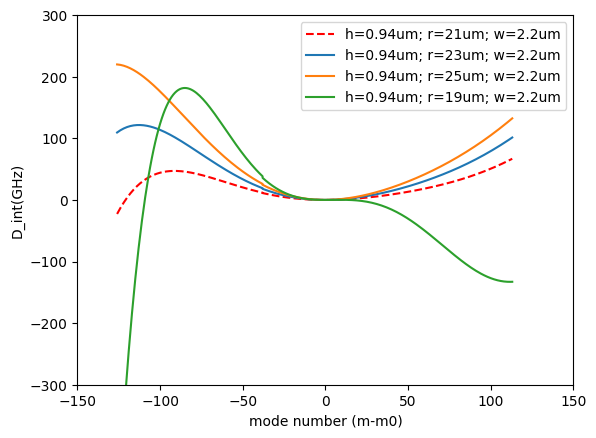}}
\caption{$D_{int}$ values of Si$_3$N$_4$ films fabricated through the LVPCD method as a function of resonator mode number.  The microring resonator designs were simulated with a height (thickness) $h = 0.94~\mu m$,  a waveguide width $w = 2.2~\mu m$ and radii values between $R = 21-29~\mu m$. The central wavelength of the resonator is 1324~nm. The variability of the parabolic behavior about the central mode will determine the width of the corresponding soliton spectrum.}
\label{figDisp}
\end{figure}

Anomalous dispersion is crucial for generating solitons inside a microresonator. Specifically, the waveguide material must exhibit parabolic positive dispersion near the pump laser wavelength in addition to higher-order dispersion terms for spectral broadening. Comb mode generation can extend into frequency regions where the integrated dispersion $D_{int}$ changes sign from positive to negative or vice versa due to the third- and fourth-order terms.  For soliton generation, a typical microresonator has an integrated dispersion ($D_{int}$) profile resembling a saddle shape, as shown in Figure \ref{figDisp}. The integrated dispersion, in frequency units, is given by $D_{int} = \omega_m - (\omega_0 + D_{1m})$, representing the influence of higher-order dispersion terms. The values for these higher-order parameters can be extracted through curve fitting to $D_{int}$. Note that group velocity dispersion (GVD) is characterized by the second derivative of $D_{int}$ and is considered anomalous when this quantity is greater than zero.

Higher-order dispersion terms have specific effects on soliton generation. For example, fourth-order dispersion affects comb bandwidth, while third-order dispersion contributes to frequency comb stability. Additionally, these higher-order terms can lead to asymmetric comb spectra and cause the soliton to emit a dispersive wave, which can expand the comb’s spectral bandwidth to two-thirds of an octave or even a full octave.

Fig. \ref{figDisp} shows the integrated dispersion of microring resonators with radii ranging from $R=21-29~\mu m$. These resonators were modeled using refractive index values based on in-house fabrication. The modeling results indicate that, in general, the slope of the quadratic dispersion increases with the ring resonator radius. The magnitude of the higher-order dispersion terms ($D_3$, $D_4$, etc.) decreases as the radius increases, with more shallow curves corresponding to broader spectra. The exception is the case for $R=29~\mu m$, where the effects of higher-order dispersion are more pronounced.

\section{Microring resonator optimization}

Optimization of combs and solitons produced by a microring resonator depends on the specific application and the desired level of complexity. Generally, combs with large free spectral ranges (FSRs $\sim10-50$ THz) are preferred due to their high tunability, low power consumption, compact size, and compatibility with other devices.

Solitons propagating through the resonator can be classified as either dissipative or breathing solitons. Dissipative solitons are self-localized structures that achieve a balance between 2 pairs of optical processes: (i) material dispersion, which lengthens the temporal extent of the pulse, is offset by Kerr non-linearity, which causes temporal compression; and (ii) propagation losses, which are counterbalanced by gain from the pump. However, there is a tipping point in the system’s equilibrium, a common occurrence in non-linear dynamics, where the compensation effects are reduced under certain conditions, destabilizing the soliton. This results in oscillation of the waveform and the formation of a breathing soliton, where the soliton collapses and reappears at different times.

\begin{figure}[htbp]
\centerline{\includegraphics[width=0.5\textwidth]{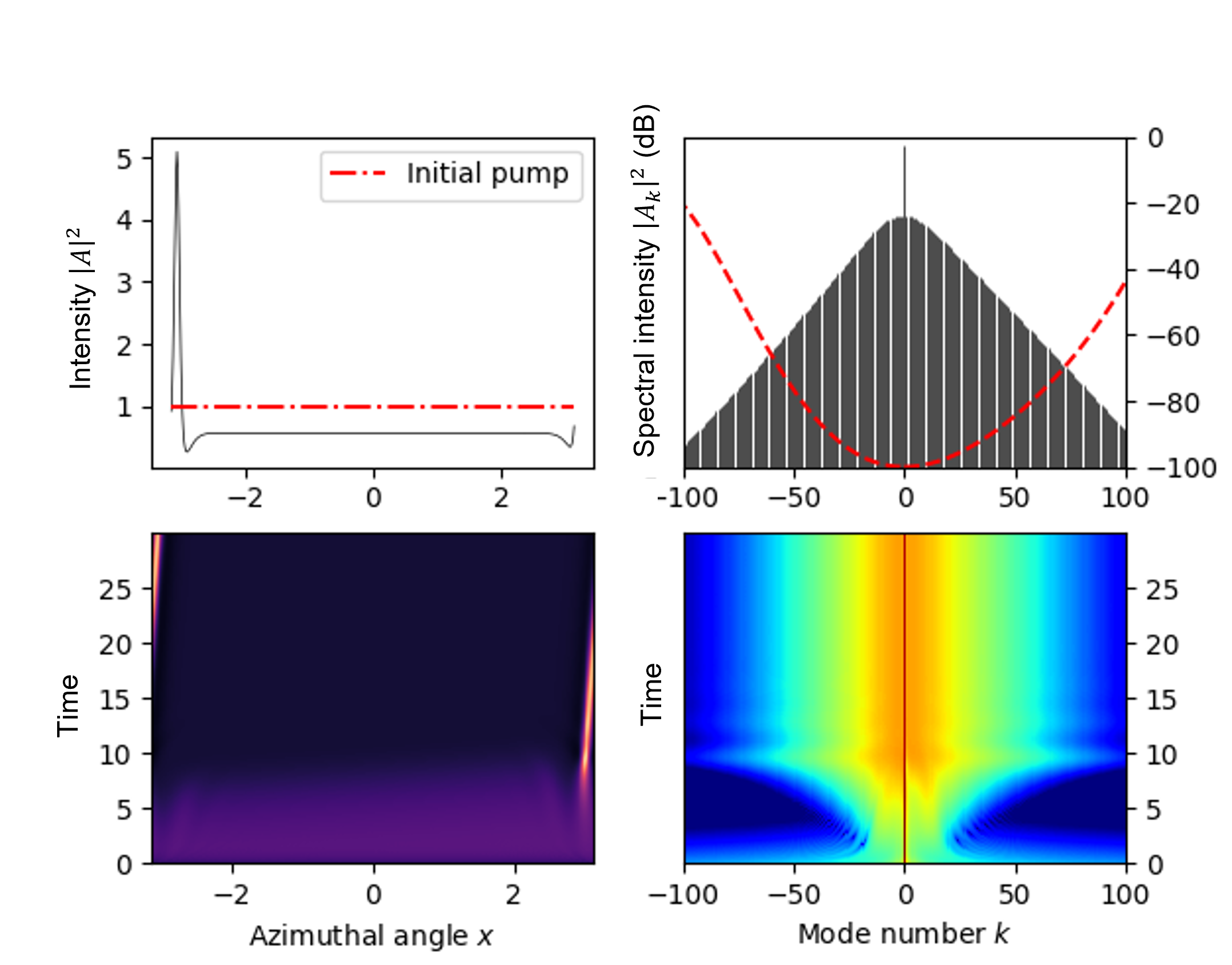}}
\caption{Single-soliton obtained from a resonator with dimensions: $w=2.2~\mu m$,
$h=0.94~\mu m$, and $R=23~\mu m$ using a CW pump field at 57~mW. The plots on the left show the input and output optical fields (top) and evolution of the field (bottom) in the time domain as a function of the azimuthal angle of the ring resonator. Plots on the right are the integrated dispersion and output soliton structure (top) and evolution of the soliton in the frequency domain (bottom) as a function of cavity mode number.}
\label{figProc}
\end{figure}

In our parameter space search, we simulated electromagnetic propagation through the microring resonator with dimensions $R = 23~\mu m$, $h = 0.94~\mu m$, and $w = 2.2~\mu m$. Figure \ref{figProc} shows the simulation results for solitons in both the time and frequency domains using the refractive index values from our in-house fabrication. Note that soliton generation depends on the input power to the microring resonator: too little power is insufficient to generate a soliton, while too much power can produce multiple solitons. In our simulations, a dissipative soliton was generated using a continuous-wave field of 57~mW at 1324~nm. Further analysis (not shown here) investigated the dependence of soliton generation on waveguide height and width while keeping the radius fixed.

Below is a summary of some of our findings:

\begin{itemize}

\item Detuning the pump from the central frequency allows for soliton generation at lower pump powers.

\item The spectral extent of a frequency comb increases with the extent of the integrated dispersion profile and the presence of higher-order dispersion.

\item Generally, the optical power required for soliton generation increases with the ring radius.

\item By varying the pump power, we can achieve soliton formation in three regimes: breathing solitons in the critical coupling regime, clean solitons in the overcoupling regime, and breathing solitons in the undercoupled regime.

\item Solitons can also be generated using a pulsed pump field, but the generation is limited to a specific range of pulse widths relative to the resonator radius.

\item For our simulations, a quality factor of $Q=10^6$ was necessary.

\item Solitons were observed for this microring resonator with pump frequencies ranging from 194 to 258 THz.

\end{itemize}

Additionally, we found that third-order or higher-order dispersion can cause some of the soliton energy to be lost due to the presence of a dispersive wave at a specific frequency. This frequency is determined by a phase-matching condition that differs from that of the four-wave mixing process. While any dispersive medium can generate such waves, their amplitude remains low in the absence of suitable phase-matching.

A summary of the optimal parameters for soliton generation with SiN microring resonators at 1324~nm is shown in Table~\ref{figParam} below.

\begin{table}[ht]
    \centering

    \begin{tabular}{|c|c|c|}
        \hline
        \textbf{Parameters} & \textbf{Values} & \textbf{Unit} \\
        \hline
        Refractive index $(n_0)$ & 1.97 & ----- \\
        \hline
        Nonlinear refractive index $(n_2)$ & $2.4 \times 10^{-19}$ & $\mathrm{m^2/W}$ \\
        \hline
        Center frequency $(\omega_0/2\pi)$ & 226 (1324 nm) & THz \\
        \hline
        Detuning $(\Delta\omega/2\pi)$ & 0.511 & GHz \\
        \hline
        FSR $(D_1/2\pi)$ & $\approx 1$ & THz \\
        \hline
        Dispersion & $(D_2/2\pi)$  124 & MHz \\
        
        & $(D_3/2\pi)$  -4.1 & kHz \\
        
        & $(D_4/2\pi)$  20 & kHz \\
        \hline
        Radius $(R)$ & 19-25 & $\mu$m \\
        \hline
        Width $(w)$ & 1.4-2.4 & $\mu$m \\
        \hline
        Height $(h)$ & 0.86-0.94 & $\mu$m \\
        \hline
        Coupling quality factor $(Q_c)$ & $10^6$ & ----- \\
        \hline
        Internal quality factor $(Q_i)$ & $10^6$ & ----- \\
        \hline
        Internal decay rate $(\kappa_i/2\pi)$ & 0.226 $\times 10^{12}$ & GHz \\
        \hline
        Coupling decay rate $(\kappa_c/2\pi)$ & 0.226 $\times 10^{12}$ & GHz \\
        \hline
        Input power $(P_{\text{in}})$ & 29-63 & mW \\
        \hline
    \end{tabular}
    \caption{Ideal simulation parameters for producing solitons in SiN microring resonators pumped at 1324~nm. The physics of the microring resonators were simulated using refractive index values of in-house fabricated SiN.}
    \label{figParam}
\end{table}

\section{Conclusion}

In conclusion, we presented a novel method of stress controlled deposition of stoichiometric silicon nitride film which is hybrid between damascene and conventional patterning processes. Specifically, we highlight the superiority of LPCVD over PECVD, owing to its ability to produce higher quality films with fewer hydrogen bond impurities at higher temperatures.

A variety of techniques were utilized to characterize films and provided insights into film composition, stress characteristics, and surface roughness. Notably, the increase in stress due to backside nitride etching resulted in a decrease in refractive index, a factor critical for achieving precise phase-matching conditions for integrated quantum optics. The final devices exhibited low loss values in-line with the present state-of-the-art. 

The refractive index values of our fabrication were implemented for the modeling of microring resonator devices and simulation of optical solitons. Optimization of the device geometries allowed us to find the ideal parameters for exhibiting a dispersion profile characteristic of soliton generation. Fabrication-to-benchtop processes that incorporate fabrication, modeling and device testing, like those presented here, will serve to further advance the development of silicon nitride-based devices for quantum communication and computational applications.

\section*{Acknowledgment}

The authors gratefully acknowledge Robert Vandusen, Angela Mccormick, Rodney Aiton and Simona Moisa for their contributions to this work.  This work was supported by the National Research Council of Canada High-Throughput Secure Networks Program and Quantum Sensors Program.

\bibliographystyle{IEEEtran}
\bibliography{IEEEQCE24Bibliography}

\end{document}